\documentclass[11pt,a4paper,onecolumn]{revtex4-2}
\usepackage[utf8]{inputenc}

\usepackage{bbm}
\usepackage{lmodern}
\usepackage{enumitem}
\usepackage{hyperref}

\usepackage{siunitx}
\usepackage{amsmath}
\usepackage{amssymb}
\usepackage{mathtools}
\usepackage{graphicx}
\usepackage[dvipsnames]{xcolor}

\DeclareMathOperator\erf{erf}

\newcommand{\ket}[1]{|#1\rangle} 

\newcommand{\ve}{\boldsymbol{e}} 
\newcommand{\vq}{\boldsymbol{q}}

\newcommand{\vk}{\boldsymbol{k}} 
\newcommand{\vx}{\boldsymbol{x}} 
\newcommand{\vA}{\boldsymbol{A}}
\newcommand{\vB}{\boldsymbol{B}}

\newcommand{\vL}{\boldsymbol{L}}

\newcommand{\vR}{\boldsymbol{R}} 
\newcommand{\vG}{\boldsymbol{G}}
\newcommand{\vC}{\boldsymbol{C}}
\newcommand{\vS}{\boldsymbol{S}}

\newcommand{\ee}{\hat{\ve}}
\newcommand{\qq}{\hat{\vq}}
\begin{document}
\title{
	Linear Scaling Approach for Optical Excitations Using Maximally Localized Wannier Functions
}

\author{Konrad Merkel}
\author{Frank Ortmann}
\affiliation{TUM School of Natural Sciences, Technical University of Munich, Germany}
\affiliation{Center for Advancing Electronics Dresden, Technische Universität Dresden, Germany}
\email{frank.ortmann@tum.de}

\begin{abstract}
We present a theoretical method for calculating optical absorption spectra based on maximally localized Wannier functions, which is suitable for large periodic systems. For this purpose, we calculate the exciton Hamiltonian, which determines the Bethe-Salpeter equation for the macroscopic polarization function and optical absorption characteristics. The Wannier functions are specific to each material and provide a minimal and therefore computationally convenient basis. Furthermore, their strong localization greatly improves the computational performance in two ways: first, the resulting Hamiltonian becomes very sparse and, second, the electron-hole interaction terms can be evaluated efficiently in real space, where large electron-hole distances are handled by a multipole expansion.
For the calculation of optical spectra we employ the sparse exciton Hamiltonian in a time-domain approach, which scales linearly with system size.  
We demonstrate the method for bulk silicon - one of the most frequently studied benchmark systems - and envision calculating optical properties of systems with much larger and more complex unit cells, which are presently computationally prohibitive.
\end{abstract}

\maketitle

\section{Introduction}

Simulations of optical properties such as UV-vis-NIR absorption or reflection spectra are crucial for designing or improving opto-electronic devices with novel materials. In this context, accurate theoretical predictions help to find suitable materials much faster and at lower cost, thus complementing and guiding experimental efforts.
However, calculating optical properties is computationally demanding, which limits calculations to small systems with only a few atoms per unit cell. The reason is that optical properties are inherently affected by many-body effects. For example, the optical response of semiconductors and insulators is determined by the Coulomb interaction between electrons and holes in a material, which leads to the formation of bound electron-hole states called excitons \cite{Frenkel1931, Knox1983, Bechstedt2015}. For the calculation of optical properties such as UV-vis-NIR absorption spectra it is therefore necessary to describe two-particle states of electrons and holes that are created upon optical excitation. A suitable description of such many-body effects can be derived in terms of a Bethe-Salpeter equation (BSE) \cite{Salpeter1951, Sham1966,Hanke1980,Strinati1984, Strinati1988, Onida2002, Bechstedt2015, Blase2020} for the polarization function. For almost all real materials, however this BSE is too difficult to solve. Important simplifications can be obtained for non-spin-polarized systems, where the BSE splits into singlet and triplet parts, which can be treated independently\cite{Bechstedt2015}. Optical transitions, described by transition matrix elements that are diagonal in spin space, cannot induce spin-flips, and it is sufficient to calculate the singlet case only, which is already a huge simplification. Furthermore, the singlet-BSE can be rewritten into a generalized eigenvalue problem and further simplified by performing the Tamm-Dancoff approximation for electronically gapped systems \cite{Benedict1998, Rohlfing2000, Bechstedt2015}. The resulting Hamiltonian matrix is still very large and dense but can in principle be diagonalized for small system sizes using popular simulation packages\cite{Puschnig2002, Gajdos2006, Yambo2009, BerkeleyGW2012}. In addition, very dense $\vk$-meshes are needed in order to obtain converged results, a problem that is known from the independent particle picture \cite{Okano2020} and which becomes more severe for excitons. This has lead to strategies like the use of hybrid meshes \cite{Fuchs2008, Schleife2009}, where specific parts of the Brillouin zone are sampled with higher precision.
Despite all these works on different computational aspects, it is still challenging to include exciton effects in the calculation of optical absorption spectra, in particular for systems with many atoms per unit cell.

In this paper we present an approach based on maximally localized Wannier functions (MLWF)\cite{Marzari1997, Marzari2012}, which can deal with large and/or complex systems.
MLWF are directly obtained from underlying quasi-particle wave functions and represent a minimal basis set that is adapted to the specific material. Moreover, they can be obtained for specific bands, e.g., near the band gap, making the calculation independent of the number of atoms in a unit cell. 
 Furthermore, we show that the resulting representation has important computational advantages, namely that the Hamiltonian matrix becomes very sparse, and can therefore be solved very efficiently, thus enabling optical calculations of large systems.
 For convenience, we use the term LSWO (linear scaling Wannier optics) for the presentation of the entire approach.

\section{Theory: Optical properties and exciton Hamiltonian}
\subsection{General formalism} \label{section:general_formalism}

We start from the two-particle eigenvalue problem in Tamm-Dancoff approximation\cite{Benedict1998, Rohlfing2000, Bechstedt2015},
\begin{align} \label{eq:excition_hamiltonian_k}
\sum_{v' c' \vk'} H_{c v\vk, \,c' v'\vk'} A^\Lambda_{c'v'\vk'} = E^\Lambda A^\Lambda_{c v\vk},
\end{align}
where $c$ and $v$ label the conduction and valence bands, respectively, and $A$ describes the exciton amplitude. The crystal momentum $\vk$ is the same for electron and hole because only vertical excitations are considered in the optical limit. The hermitian singlet-exciton Hamiltonian $H$ is given by
\begin{align} \label{eq:exciton_hamiltonian_k}
H_{c v\vk, \,c' v'\vk'} =& \left[ E^\text{cond.}_c(\vk) - E^\text{val.}_v(\vk) \right] \delta_{cc'} \delta_{vv'} \delta_{\vk\vk'} - H^\text{SC}_{cv \vk,\, c'v' \vk'} + 2 H^\text{LFE}_{cv \vk,\, c'v' \vk'}
\end{align}
and consists of effective single-particle contributions from conduction and valence band structures (first term), which are diagonal with respect to $\vk$, and two-particle contributions from screened electron-hole interactions $H^\text{SC}$ and local field effects $H^\text{LFE}$, which couple different $\vk$ and $\vk'$ via Coulomb interaction.
While the occurrence of a screened electron-hole interaction is intuitively plausible, the local field effects (LFE) term seems less obvious and some comments are appropriate. LFE arise when the system is inhomogeneous on the microscopic scale, i.e. the microscopic dielectric function $\epsilon_{\vG\vG'}$ is not diagonal with respect to reciprocal lattice vectors $\vG$\cite{Wiser1963, Hanke1975, Gavrilenko1996}. By including LFE in the Hamiltonian, it is ensured that one can later calculate the macroscopic rather than the microscopic dielectric tensor directly from $E^\Lambda$ and $A^\Lambda$. Note that the LFE matrix elements are in the form of electron-hole pair exchange interactions.\cite{Combescot2023}

$H^\text{SC}$ and $H^\text{LFE}$ can be obtained from single-particle Bloch functions for conduction $\phi_{c\vk}(\vx)$ and valence states $\phi_{v\vk}(\vx)$. A natural choice for $\phi_{c\vk}(\vx)$ and $\phi_{v\vk}(\vx)$ are Kohn-Sham orbitals leading to
\begin{align} \label{eq:screenedCoulomb_k}
H^\text{SC}_{cv \vk, \,c'v' \vk'} =& \int dx \int dx' \, \phi^*_{c\vk}(\vx) \phi^*_{v'\vk'}(\vx') W(\vx-\vx') \phi_{v\vk}(\vx') \phi_{c'\vk'}(\vx), \\
H^\text{LFE}_{cv \vk,\, c'v' \vk'} =& \int dx \int dx' \,  \phi^*_{c\vk}(\vx) \phi^*_{v'\vk'}(\vx')\left[\frac{1}{\Omega} \sum_{\vG\ne 0} \tilde{V}(|\vG|) e^{i\vG(\vx-\vx')} \right]  \phi_{v\vk}(\vx) \phi_{c'\vk'}(\vx'), \label{eq:LFE_k}
\end{align}
where $W(\vx-\vx')$ is the screened Coulomb interaction and 
$\tilde{V}(|\vq+\vG|) =  \frac{4\pi e^2}{\epsilon_0} \frac{1}{|\vq+\vG|^2}$ is the Fourier transformed bare Coulomb potential. The screening might be obtained from different approaches, including a GW calculation or using a model screening function or just a constant relative permittivity.
Here, we use a model dielectric function $\epsilon^{-1}(\vq) = 1- (\eta + \alpha q^2 / q_\text{TF}^2)^{-1}$ that has been shown to yield good results for typical semiconductors \cite{Cappellini1993}. The parameter $\eta = (1- \epsilon_\infty^{-1})^{-1}$ with the electronic dielectric constant $\epsilon_{\infty}$ of the material, and $q_{\text{TF}}$ is the Thomas-Fermi wave vector. The dimensionless parameter $\alpha=1.563$ has been shown to be rather universal \cite{Cappellini1993}. The screened Coulomb potential is then obtained from $W(\vq) = \epsilon^{-1}(\vq) \tilde{V}(\vq)$. We assume a static screening, i.e. no time dependence, which is the most frequent approach. However, we note that current efforts also investigate extensions to the frequency dependence of screening \cite{Zhang2023,Sangalli2023}.
By taking the Fourier transform we obtain the corresponding potential in real space,
\begin{align} \label{eq:model_screenning_realspace}
W(\vx-\vx') &= \frac{1}{4\pi \epsilon_{0} \epsilon_\infty |\vx-\vx'|} + \left(1- \epsilon_\infty^{-1} \right)\frac{\exp\left[ \frac{- q_\text{TF}  |\vx-\vx'|}{ \sqrt{(1- \epsilon_\infty^{-1})\alpha}} \right]}{4\pi \epsilon_{0} |\vx-\vx'|} \nonumber \\
&= V_{\text{scr}}(|\vx-\vx'|) + \left(1-\epsilon_\infty^{-1} \right) V_{\text{Yuk}}(|\vx-\vx'|),
\end{align}
which is the superposition of a screened Coulomb and a Yukawa potential. A more detailed derivation can be found in Section~\ref{sec:model_screening_potential} of the appendix.

Independently of the type of screening, the numerical evaluation of Eq.~\eqref{eq:excition_hamiltonian_k} can be quite expensive because a very fine $\vk$-mesh is usually required to obtain converged results and the Hamiltonian matrix that needs to be diagonalized is very large and, in general, a dense matrix.
Furthermore, the underlying Bloch functions, that are needed for the evaluation of Eq.~\eqref{eq:screenedCoulomb_k} and Eq.~\eqref{eq:LFE_k}, are delocalized which leads to additional challenges for numerical calculations. These obstacles are circumvented by transforming above equations into a localized basis of Wannier functions which will be explained here below.

\subsection{Exciton-Hamiltonian in basis of MLWF}

For an efficient treatment of the exciton problem in Eq.~\eqref{eq:excition_hamiltonian_k}, it is advantageous to employ a localized basis of MLWF $w_{m\vR}(\vx)$. 
MLWF are routinely used to investigate single-particle observables \cite{Marzari2012, Merkel2023} and have been shown to be advantageous for many-body first-principles calculations, including electron-electron interactions and screening\cite{Koshino2018, Shih2012}, spin excitations\cite{Schindlmayr2010} or quadratic optical response\cite{Garcia2023}.
They are directly related to the underlying Bloch functions $\phi_{n\vk}(\vx)$ by the transformation,
\begin{align}
w_{m\vR}(\vx) &:= \frac{1}{\sqrt{N_\Omega}} \sum_{n\vk} e^{-i\vk \vR} U_{mn}(\vk) \phi_{n\vk}(\vx),
\end{align}
where $\vR$ represents a unit cell vector and $U(\vk)$ is a unitary matrix. It can be chosen such that the obtained Wannier functions are maximally localized, i.e. their spread $\left[ \langle \vx^2 \rangle - {\langle\vx\rangle}^2 \right]$ is minimal. To be more precise, $U(\vk)$ disentangles the individual energy bands in case of band crossings or degeneracies and fixes the $\vk$-dependent gauge phase $e^{i\theta(\vk)}$ that each Bloch function has. $U(\vk)$ can be obtained from an optimization algorithm\cite{Marzari1997, Marzari2012} for specific groups of bands, e.g. all valence bands.
The obtained MLWF are orthogonal to each other and must be real valued\cite{Marzari1997}.
Owing to translational symmetry, MLWF at different unit cells $\vR$ have the same shape and are related to each other by $w_{m\vR}(\vx) = w_{m0}(\vx-\vR)$ , which is known as shift property. 

For the LSWO approach it is advantageous to obtain MLWF for conduction and valence bands near the fundamental band gap separately. Therefore, the obtained MLWF keep the character of either an electron or a hole. We denote them as conduction-WF and valence-WF in the following. Even though the conduction and valence MLWF are obtained separately, they are orthogonal since valence and conduction states are non-degenerate for all $\vk$-points. Hence, they represent a suitable basis for the excitonic two-particle Hilbert space.

As mentioned above, only a subspace of the two-particle Hilbert space in which electrons and holes have the same momentum is relevant for the calculation of optical properties. This means we need to transform the Bloch representation with the indexes $cv\vk$ into a real-space description of MLWF with indexes $mn\vS$. This mapping is achieved by a unitary transformation of the two particle basis using the matrix
\begin{align}
F_{mn\vS, \,cv\vk} &= \frac{1}{\sqrt{N_\Omega}} e^{i\vk\vS} U^*_{cm}(\vk) U_{nv}(\vk),
\end{align}
where the $U$ matrices are obtained from Wannier transformations of valence and conduction bands and the unit cell vector $\vS=\vR-\vL$ is the distance between electron unit cell $\vR$ and hole unit cell $\vL$.
Excitonic wave functions in the optical subspace (i.e. at vanishing photon momentum $\vq\to 0$) are obtained by
\begin{align}
\xi_{mn\vS}(\vx,\vx')
&=\sum_{cv\vk} F_{mn\vS, \,cv\vk}\,  \phi^*_{c\vk}(\vx)\phi_{v\vk}(\vx') \nonumber\\
&= \frac{1}{\sqrt{N_\Omega}} \sum_{\vR} w_{m\vR}(\vx) w_{n,\vR-\vS}(\vx').
\label{eq:xi_wannier}
\end{align}
We have used that MLWF are real and therefore the excitonic wave function fulfills $\xi_{mn\vS}=\xi_{mn\vS}^*$. Eq.~\eqref{eq:xi_wannier} is a manifestation of the convolution theorem in terms of Bloch functions and corresponding MLWF. At this point we should mention that the use of the variable $\vR$ (electron unit cell) as summation index by no means introduces any asymmetry in the treatment of electrons and holes.
The same result can also be expressed by centre of mass and relative coordinates. The centre of mass motion is not relevant for optics due to translational symmetry of the crystal and only the relative distance $\vS$ between electron and hole remains in $\xi_{mn\vS}$.

We also use $F_{mn\vS, \,cv\vk}$ to transform Eq.~\eqref{eq:excition_hamiltonian_k} into the Wannier basis,
\begin{align} \label{eq:excition_hamiltonian_wannier}
\sum_{m'n'\vS'} \tilde{H}_{mn\vS, \,m'n'\vS'} B^\Lambda_{ m'n'\vS'} &= E^\Lambda B^\Lambda_{mn\vS},
\end{align}
where the exciton eigenvector is obtained as
\begin{align}
B^\Lambda_{mn\vS} = \sum_{cv\vk} F_{mn\vS,\,cv\vk}\, A^\Lambda_{c v\vk}
\end{align}
and the exciton Hamiltonian becomes
\begin{align}
\tilde{H}_{mn\vS, \,m'n'\vS'} =& \sum_{cv\vk} \sum_{c'v'\vk'} F_{mn\vS, \,cv\vk}\, H_{c v\vk, \,c' v'\vk'}\, F_{c'v'\vk', \,m'n'\vS'}^* \nonumber \\
 =& \tilde{H}^\text{band}_{mn\vS,\, m'n'\vS'} - \tilde{H}^\text{SC}_{mn \vS,\, m'n' \vS'} + 2 \tilde{H}^\text{LFE}_{mn \vS, \,m'n' \vS'}.
 \label{eq:exciton_Hamiltonian_wannier}
\end{align}
According to Eq.~\eqref{eq:exciton_hamiltonian_k} the single-particle band contributions are obtained as
\begin{align} \label{eq:single_particle_H_wannier}
\tilde{H}^\text{band}_{mn\vS,\, m'n'\vS'} 
&= H^\text{cond.}_{m'm}(\vS-\vS') \delta_{nn'} - H^\text{val.}_{nn'}(\vS-\vS') \delta_{mm'},
\end{align}
where $H^\text{cond.}_{m'm}(\vS-\vS')$ and $H^\text{val.}_{nn'}(\vS-\vS')$ are the single-particle Wannier Hamiltonians for conduction and valence bands, respectively. They are directly accessible from the Wannier transformation of the first-principles electronic structure. \cite{Marzari1997, Marzari2012}

The screened electron-hole interaction can be obtained by virtue of Eq.~\eqref{eq:xi_wannier} and by applying the shift property of MLWF (see appendix),
\begin{align}
\tilde{H}^\text{SC}_{mn \vS,\, m'n' \vS'} &= \int dx \int dx' \, \xi_{mn\vS}(\vx,\vx')   W(\vx-\vx') \xi_{m'n'\vS'}(\vx,\vx') \nonumber\\
&= \sum_{\vA} \tilde{W}^{mm'}_{nn'}(\vA, \vS, \vS'), \label{eq:Coulomb_integral_wannier}
\end{align}
with the general Coulomb matrix elements
\begin{align} \label{eq:Coulomb_integral_wannier_single}
\tilde{W}^{mm'}_{nn'}(\vA, \vS, \vS') &= \int dx \int dx' \, 
w_{m0}(\vx) w_{m'\vA}(\vx)   W(\vx-\vx')  w_{n',\vA-\vS'}(\vx') w_{n,-\vS}(\vx') \nonumber \\
&=  \tilde{W}^{m'm}_{n'n}(-\vA, \vS', \vS) ,
\end{align}
which depend on three different unit cell vectors (corresponding to three $\vk$-vectors in reciprocal space). $\tilde{H}^\text{SC}_{mn \vS, \,m'n' \vS'}$ only depends on two unit cell vectors because electrons and holes have the same momentum.
For a more intuitive and physically comprehensible description, we introduce the unit cell vectors $\vR_c$, $\vR_v$, and $\vR_D$, which correspond to the relative shifts between conduction WFs, between valence WFs, and to the electron-hole distance, respectively. We substitute  $\vA = \vR_c$, $\vS=-\vR_D$ and $\vS' = -\vR_D+\vR_c-\vR_v$ in Eq.~\eqref{eq:Coulomb_integral_wannier_single} and use the shift property of MLWF to obtain
\begin{align} \label{eq:Coulomb_integral}
\tilde{W}^{mm'}_{nn'}&(\vA = \vR_c,\vS=-\vR_D, \vS'=-\vR_D+\vR_c-\vR_v) =\\\nonumber
 &= W^{mm'}_{nn'}(\vR_c,\vR_v, \vR_D) = \int d^3 x  \int d^3 x' \rho_{mm'\vR_c}(\vx) W(\vx-\vx'-\vR_D) \rho_{nn'\vR_v}(\vx'),
\end{align}
where $\rho_{mm'\vR_c}(\vx) = w_{m0}(\vx) w_{m'\vR_c}(\vx)$ and $\rho_{nn'\vR_v}(\vx) = w_{n0}(\vx) w_{n'\vR_v}(\vx)$ are (overlap) densities of two electrons and (overlap) densities of two holes, respectively.

Before we come to the integration strategy in Sect. \ref{sec:numerical_impl}, we comment on the distance dependence of these matrix elements.
Since the overlap between two different MLWF is exponentially suppressed with increasing distance, it is clear that the overlap densities vanish for large values of $\vR_c$ and $\vR_v$. 
Therefore, the corresponding Coulomb integrals Eq.~\eqref{eq:Coulomb_integral} also vanish rapidly for large displacements $\vR_c$ or $\vR_v$. This substantially reduces the number of calculations required and constitutes a significant advantage over a plane wave basis set. In contrast, $\vR_D$ is associated with long-range Coulomb interactions, which always yields contributions that decay very slowly.
Substituting back the original variables $\vS$, $\vS'$, and $\vA$, we see that finite Coulomb integrals contribute only to matrix elements $\tilde{H}^\text{SC}_{mn \vS, \, m'n' \vS'}$ near the diagonal and $\vR_D$ corresponds to the position along the diagonal.
The matrix representation is therefore very sparse. This is a great advantage for numerical computations, since diagonalization or alternative  treatments can be performed very efficiently and with low memory requirements. It is thus not surprising that other localized basis sets leading to sparse representations of Coulomb interactions have shown large performance advantages for GW calculations in the past. \cite{Foerster2011, Wilhelm2018}
The diagonal elements for which $m=m'$, $n=n'$, and $\vR_c=\vR_v=0$ (or alternatively $\vA = 0$ and $\vS=\vS'=-\vR_D$) are expected to yield the largest contributions to $\tilde{H}^\text{SC}$. They represent interactions of classical charge densities with total charge of one, because MLWF are normalized.
The non-diagonal elements of $\tilde{H}^\text{SC}$ correspond to interactions where at least one density is an overlap density, i.e. $\rho_{mm'\vR_c}$ or $\rho_{nn'\vR_v}$ contains two different MLWF. Such overlap densities have zero total charge because MLWF are orthogonal. We therefore expect the non-diagonal elements to be significantly smaller.
Finally, contributions from LFE, Eq.~\eqref{eq:LFE_k}, are calculated in analogy to  Eq.~\eqref{eq:Coulomb_integral_wannier},
\begin{align}
\tilde{H}^\text{LFE}_{mn \vS,\, m'n' \vS'} &= \int dx \int dx' \, \xi_{mn\vS}(\vx,\vx)   \bar{V}(\vx-\vx') \xi_{m'n'\vS'}(\vx',\vx') \nonumber\\
&= \int dx \int dx' \, w_{m0}(\vx) w_{n,-\vS}(\vx) \left[\sum_{\vG\ne 0} \tilde{V}(|\vG|) e^{i\vG(\vx-\vx')} \right] w_{m'0}(\vx') w_{n',-\vS'}(\vx').
 \label{eq:LFE_wannier}
\end{align}
This matrix is, like $\tilde{H}^\text{SC}$, very sparse since the overlap between MLWF is exponentially suppressed with increasing distance. Consequently, only matrix elements with small values $\vS$ and $\vS'$, where electron and hole have closest distance, are affected by LFE. In the limiting case of strongly localized Wannier functions only matrix elements with $\vS=\vS'=0$ would contribute.
We thus have a complete description of the singlet exciton Hamiltonian in the Wannier basis Eq.~\eqref{eq:excition_hamiltonian_wannier} that can be used to calculate optical properties.

\subsection{Optical properties}

The macroscopic dielectric function $\epsilon^\text{M} (\qq, \omega)$ could be calculated within the original Bloch representation directly from the solutions of Eq.~\eqref{eq:excition_hamiltonian_k} and the optical transition matrix elements $M_{cv\vk}(\qq)$ that can be obtained from conduction and valence Bloch functions,
\begin{align} \label{eq:transition_matrix_k}
M_{cv\vk}(\qq) = \lim_{\vq\to 0} \frac{e}{\sqrt{4\pi \epsilon_{0}} |\vq|i} \int d^3x \phi^*_{c\vk}(\vx) e^{i\vq\vx} \phi_{v\vk}(\vx).
\end{align}
The macroscopic dielectric function is given as\cite{Bechstedt2015}
\begin{align} \label{eq:mac_dielec_k}
\epsilon^\text{M} (\qq, \omega) = 1+ \frac{4\pi}{\Omega} \sum_{\Lambda} \left| \sum_{cv\vk} M^*_{cv\vk}(\qq) A^\Lambda_{c v\vk} \right|^2 \left[\frac{1}{E^\Lambda-\hbar(\omega+i\eta)} + \frac{1}{E^\Lambda+\hbar(\omega+i\eta)}\right].
\end{align}
Like in the previous section we transform these expressions into the basis of MLWF by utilizing the matrix  $F_{mn\vS,\,cv\vk}$ to calculate $\epsilon^\text{M} (\qq, \omega)$ directly from $ B^\Lambda_{mn\vS}$ and corresponding transition matrix elements.
The transformation is applied to the scalar product in Eq.~\eqref{eq:mac_dielec_k},
\begin{align}
\sum_{cv\vk} M^*_{cv\vk}(\qq) A^\Lambda_{c v\vk} &=
\sum_{mn\vS} \sum_{c'v'\vk'} M^*_{c'v'\vk'}(\qq)
 F_{c'v'\vk',\,mn\vS}^* \sum_{cv\vk} F_{mn\vS,\,cv\vk} \,
A^\Lambda_{c v\vk} \nonumber \\
&=
\sum_{mn\vS} \tilde{M}^*_{mn\vS}(\qq) B^\Lambda_{mn\vS} ,
\end{align}
where $\tilde{M}^*_{mn\vS}(\qq) = \sum_{c'v'\vk'} M^*_{c'v'\vk'}(\qq) F_{c'v'\vk',\,mn\vS}^*$ was defined in the last step. Using Eq.~\eqref{eq:xi_wannier} we can rewrite the transition matrix elements in terms of MLWF,
\begin{equation}
\tilde{M}^*_{mn\vS}(\qq)
= \lim_{\vq\to 0} \frac{ie}{\sqrt{4\pi \epsilon_{0}} |\vq|} \frac{1}{\sqrt{N_\Omega}} \sum_{\vR} \int d^3x \, w_{m0}(\vx)  e^{-i\vq(\vx+\vR)} w_{n,-\vS}(\vx).
\end{equation}
Taylor expanding the exponential up to linear order (higher orders are irrelevant in the optical limit $q\to 0$) \cite{Agranovich1984,Muller2023} we get
\begin{equation}
\tilde{M}^*_{mn\vS}(\qq) = \frac{e\sqrt{N_\Omega}}{\sqrt{4\pi \epsilon_{0}}} \qq \int d^3x \, w_{m0}(\vx) \vx w_{n,-\vS}(\vx).
\label{eq:transition_wannier}
\end{equation}
From Eq.~\eqref{eq:transition_wannier} we can see that the transition matrix elements are proportional to transition dipole moments, i.e. dipole moments of electron-hole overlap densities, which nicely connects to expectations from finite systems. 
The evaluation of transition dipole moments does not cause any problems (like one would have with delocalized Bloch functions) since Wannier functions are localized in real space. 
Finally, the macroscopic dielectric function becomes
\begin{align} \label{eq:mac_dielec_wannier}
\epsilon^\text{M} (\qq, \omega) = 1+ \frac{4\pi}{\Omega} \sum_{\Lambda} \left| \sum_{mn\vS} \tilde{M}^*_{mn\vS}(\qq) B^\Lambda_{mn\vS} \right|^2 \left[\frac{1}{E^\Lambda-\hbar(\omega+i\eta)} + \frac{1}{E^\Lambda+\hbar(\omega+i\eta)}\right].
\end{align}
With Eqs.~\eqref{eq:mac_dielec_wannier},\eqref{eq:transition_wannier} and \eqref{eq:exciton_Hamiltonian_wannier} the entire problem is formulated in the Wannier basis. The remaining task is to evaluate all required matrix elements for the screened Coulomb interaction and LFE in this basis, which will be discussed below.

\section{Numerical evaluation of two-particle matrix elements and macroscopic dielectric function} \label{sec:numerical_impl}

\subsection{Evaluating Coulomb matrix elements in the basis of MLWF}  \label{sec:evaluating_coulomb}

For the numerical evaluation of the screened Coulomb interaction we insert the model-screened potential Eq.~\eqref{eq:model_screenning_realspace} into Eq.~\eqref{eq:Coulomb_integral} and evaluate the Coulomb and Yukawa potentials separately,
\begin{align}
W^{mm'}_{nn'}(\vR_c,\vR_v, \vR_D) =& \int d^3 x  \int d^3 x' \rho_{mm'\vR_c}(\vx) V_{\text{scr}}(|\vx-\vx'-\vR_D|) \rho_{nn'\vR_v}(\vx')  \nonumber\\
&+ \left(1-\epsilon_\infty^{-1} \right) \int d^3 x  \int d^3 x' \rho_{mm'\vR_c}(\vx) V_{\text{Yuk}}(|\vx-\vx'-\vR_D|) \rho_{nn'\vR_v}(\vx'). \label{eq:coulomb_yukawa_integral}
\end{align}
While the integral with the Yukawa potential (second term of Eq.~\eqref{eq:coulomb_yukawa_integral}) can be solved efficiently in reciprocal space, the numerical evaluation of the Coulomb integral (first term of Eq.~\eqref{eq:coulomb_yukawa_integral}) is quite challenging, because the potential diverges in both real and reciprocal space for $\vx\to0$ and $\vq\to0$. However, the integral is nevertheless finite as can be shown on general grounds.
The problem is still complicated by the fact that MLWF are typically obtained numerically from DFT or GW calculations and analytic forms are usually unknown.
Strategies to circumvent such issues include expansions of MLWF using spherical harmonics and appropriate radial functions \cite{Schnell2002, Schnell2003}, where the Coulomb integrals can be rewritten and partly solved analytically, or attempts to expand MLWF around the origin in $\vk$-space by a suitable Taylor expansion.
While the latter is numerically inconvenient, the expansion in spherical harmonics can provide satisfactory results for simple systems \cite{Schnell2002}, especially when the Wannier functions are expressed in a form of atomic orbitals and only a small number of expansion coefficients are needed. This, however, may not be the case, which means that in general an extreme large set of spherical harmonics becomes necessary, especially when satellite structures far away from the charge centre exist.
Alternatively, one might consider choosing a different system of functions where the Coulomb integrals can be solved analytically. A well-known example is Gaussian basis functions, which are routinely used in quantum chemistry codes\cite{Gill1994}.
However, an expansion of MLWF in terms of such basis functions is usually very complicated and requires sophisticated optimization and fitting algorithms. Despite some proof of principle studies \cite{Bakhta2018}, there are no commonly available tools to perform such an elaborated task.
Here, we want to use a numerical method that yields satisfactory results for all types of MLWF and is easily applicable. This method follows the ab-initio philosophy in the sense that we avoid any fitting.

The numerical evaluation of the first term of  Eq.~\eqref{eq:coulomb_yukawa_integral} is performed in multiple steps. We start by introducing auxiliary densities $\rho^\text{aux}_{mm'\vR_c}(\vx)$ and $\rho^\text{aux}_{nn'\vR_v}(\vx)$  for each $\rho_{mm'\vR_c}(\vx)$ and $\rho_{nn'\vR_v}(\vx)$, respectively.
These auxiliary densities are Gaussian functions with the constraint that they have the same charge as the corresponding overlap density, i.e.,
\begin{align} \label{eq:auxiliary_gauss_norm}
\int d^3 x \,\rho^\text{aux}_{mm'\vR_c}(\vx) = \int d^3 x \,\rho_{mm'\vR_c}(\vx).
\end{align}
The centre and variance of each Gaussian function is in general not important, albeit specific choices might be numerically favourable.
We continue by adding and subtracting auxiliary densities for each integral and separate four different terms,
\begin{align}
&\int d^3 x \int d^3 x' \left[\rho_{mm'\vR_c}(\vx) - \rho^\text{aux}_{mm'\vR_c}(\vx) + \rho^\text{aux}_{mm'\vR_c}(\vx)  \right] \times \nonumber \\ &\times V_\text{scr}( \vx-\vx' -\vR_D ) 
 \left[\rho_{nn'\vR_v}(\vx')- \rho^\text{aux}_{nn'\vR_v}(\vx') + \rho^\text{aux}_{nn'\vR_v}(\vx') \right] \nonumber \\
=& I_1 + I_2 + I_3 + I_4,
\end{align}
where the individual contributions are given by,
\begin{align}
I_1 =& \int d^3 x \int d^3 x' [\rho_{mm'\vR_c}(\vx) - \rho^\text{aux}_{mm'\vR_c}(\vx)] V_\text{scr}( \vx-\vx'-\vR_D  ) [\rho_{nn'\vR_v}(\vx')- \rho^\text{aux}_{nn'\vR_v}(\vx')] , \nonumber\\
I_2 =& \int d^3 x \int d^3 x' [\rho_{mm'\vR_c}(\vx) - \rho^\text{aux}_{mm'\vR_c}(\vx)] V_\text{scr}( \vx-\vx'-\vR_D  ) \rho^\text{aux}_{nn'\vR_v}(\vx') , \nonumber\\ 
I_3 =& \int d^3 x \int d^3 x' \rho^\text{aux}_{mm'\vR_c}(\vx) V_\text{scr}( \vx-\vx' -\vR_D ) [\rho_{nn'\vR_v}(\vx')- \rho^\text{aux}_{nn'\vR_v}(\vx')], \nonumber \\
I_4 =&\int d^3 x \int d^3 x' \rho^\text{aux}_{mm'\vR_c}(\vx) V_\text{scr}( \vx-\vx'-\vR_D  ) \rho^\text{aux}_{nn'\vR_v}(\vx').
\end{align}
The last term $I_4$ can be evaluated analytically because only Gaussian functions are involved. For instance, chosing radial symmetrical Gaussians
$\rho^\text{aux}_{mm'\vR_c}(\vx) = \left(\frac{\alpha}{\pi}\right)^{3/2} e^{-\alpha |\vx - \vB|^2}$ and $\rho^\text{aux}_{nn'\vR_v}(\vx) = \left(\frac{\gamma}{\pi}\right)^{3/2} e^{-\gamma |\vx - \vC|^2}$, one obtains\cite{Gill1994},
\begin{align} \label{eq:gaussian_analytic}
I_4 =& \frac{1}{\epsilon_0\epsilon_\infty|\vB-\vC-\vR_D|} \,\erf\left[\sqrt{\frac{\alpha\gamma}{\alpha+\gamma}} |\vB-\vC-\vR_D|\right].
\end{align}
The remaining three terms $I_1$, $I_2$ and $I_3$ are solved in Fourier space. This is demonstrated for $I_1$, which, in Fourier space reads
\begin{align} \label{eq:I1_fourier}
I_1 = \frac{1}{(2\pi)^3} \int d^3q \, e^{i\vq \vR_D} f_{mm'\vR_c}(\vq) \tilde{V}_\text{scr}(\vq) f_{nn'\vR_v}(-\vq),
\end{align}
where the Fourier transformed quantities are
\begin{align}
f_{mm'\vR_c}(\vq) &= \int d^3x \, e^{-i\vq \vx} [\rho_{mm'\vR_c}(\vx) - \rho^\text{aux}_{mm'\vR_c}(\vx)], \label{eq:FT_fc}\\
f_{nn'\vR_v}(\vq) &= \int d^3x \, e^{-i\vq \vx} [\rho_{nn'\vR_v}(\vx)- \rho^\text{aux}_{nn'\vR_v}(\vx)] \label{eq:FT_fv}
\end{align}
and the Fourier transformed potential $\tilde{V}_\text{scr}(\vq)\propto q^{-2}$. The divergence at $\vq\to 0$ is integrable, i.e. the integral is finite for all finite regions including volumes around the origin.  

Since the auxiliary densities have the same charge as the corresponding overlap densities  (cf. Eq.~\eqref{eq:auxiliary_gauss_norm}), it becomes clear that $f_{mm'\vR_c}(\vq=0) = f_{nn'\vR_v}(\vq=0) = 0$ by construction. For a discrete numerical evaluation of the integral Eq.~\eqref{eq:I1_fourier}, this means that the $\vq=0$ term can be omitted, since it must be zero (finite value times zero). The only remaining task is to perform the $\vq$-sum for all $\vq\ne 0$, where no problems occur, and we obtain
\begin{align}
I_1 \simeq \frac{ \Delta V_q}{N_\text{grid}} \sum_{\vq\ne 0} \, e^{i\vq \vR_D} f_{mm'\vR_c}(\vq) \tilde{V}_\text{scr}(\vq) f_{nn'\vR_v}(-\vq).
\end{align}
Integrals $I_2$ and $I_3$ are solved in full analogy. After summation and back substitution we obtain the desired (screened) Coulomb matrix elements Eq.~\eqref{eq:Coulomb_integral_wannier_single}.

\subsection{Evaluating LFE in the basis of MLWF} \label{sec:evaluating_lfe}

The numerical calculation of LFE matrix elements in Eq.~\eqref{eq:LFE_wannier} is much easier than the screened Coulomb interaction because the used potential is not divergent ($\vG=0$ is not contained). The potential in Fourier space is obtained as,
\begin{align}
\tilde{\bar{V}}(\vq) &= \int d^3x \, e^{-i\vq \vx} \sum_{\vG\ne 0} \tilde{V}(|\vG|) e^{i\vG\vx} = (2\pi)^3 \sum_{\vG\ne 0} \tilde{V}(|\vG|) \delta(\vq-\vG).
\end{align}
The overlap densities are now between conduction and valence WF and are known as transition densities. We denote their Fourier transform as
\begin{align}
f_{mn-\vS}(\vq) &= \int d^3x \, e^{-i\vq \vx} \rho_{mn-\vS}(\vx). \label{eq:FT_lfe}
\end{align}
Finally, Eq.~\eqref{eq:LFE_wannier} becomes
\begin{equation}
\tilde{H}^\text{LFE}_{mn \vS, \,m'n' \vS'} =
\sum_{\vG\ne 0} f_{mn-\vS}(\vG) \tilde{V}(|\vG|) f_{m'n'-\vS'}(-\vG),
\label{eq:LFE_wannier2}
\end{equation}
which can be easily evaluated numerically with a Fast Fourier algorithm.

\subsection{Time domain approach for calculating the macroscopic dielectric function} \label{sec:time_domain_approach}

We have now everything at hand to construct the exciton Hamiltonian in the basis of MLWF. The remaining task would be to solve the eigenvalue equation and use Eq.~\eqref{eq:mac_dielec_wannier} to obtain the macroscopic dielectric function $\epsilon^\text{M}$. Numerically this could be done by using a sparse matrix diagonalization algorithm. However, we want to use a time-domain approach\cite{Schmidt2003} which allows us to calculate $\epsilon^\text{M}$ without a formal high-scaling diagonalization or restrictions to a few number of eigenvalues.
Therefore, we rewrite Eq.~\eqref{eq:mac_dielec_wannier} in the time domain by taking a Fourier transform. We start with the dielectric function in the Cartesian direction $\ee_j$,
\begin{align}
\epsilon^\text{M}_{jj} (\omega) = 1+ \frac{4\pi}{\Omega} \sum_{\Lambda} \left| \sum_{mn\vS} \tilde{M}^*_{mn\vS}(\ee_j) B^\Lambda_{mn\vS} \right|^2 \left[\frac{1}{E^\Lambda-\hbar(\omega+i\eta)} + \frac{1}{E^\Lambda+\hbar(\omega+i\eta)}\right]
\end{align}
This is equivalent to a time-domain formulation \cite{Schmidt2003},
\begin{align} \label{eq:time_approach}
\epsilon^\text{M}_{jj} (\omega) = 1- \frac{8\pi}{\Omega \hbar} \int_0^\infty dt \, e^{i(\omega+i\eta)t} \, \text{Im} \left[ \sum_{mn\vS} \tilde{M}^*_{mn\vS}(\ee_j) \psi^{(j)}_{mn\vS}(t) \right],
\end{align}
where the time-initial state is given by $\psi^{(j)}_{mn\vS}(t=0)= \tilde{M}^*_{mn\vS}(\ee_j)$ and is propagated with the exciton Hamiltonian,
\begin{align} \label{eq:time_evolution}
\psi^{(j)}_{mn\vS}(t) &= \sum_{m'n'\vS'} \left(\exp \left[\frac{-it}{\hbar} \tilde{H} \right]\right)_{mn\vS,\,m'n'\vS'} \psi^{(j)}_{m'n'\vS'}(t=0).
\end{align}

\section{Computational details} \label{sec:computational_details}

To demonstrate our approach for the example of silicon crystals, which has been frequently studied experimentally and theoretically in the past,\cite{Benedict1998, Arnaud2001, Puschnig2002, Schmidt2003} we proceed in multiple steps. First, electronic states are obtained using density functional theory (DFT) with the PBE exchange-correlation functional and PAW pseudo potentials\cite{Blochl1994, kresse1999} as implemented in the \textsc{vasp} code\cite{Kresse1996, Kresse1996_2}. We use an energy cut-off of \SI{350}{\eV} and a $11\times11\times11$ Monkhost-Pack $\vk$-points grid for converged DFT calculations.
From these results, we calculate four MLWF which correspond to all valence bands and six MLWF for the lowest-energy conduction bands separately by utilizing the \textsc{wannier90} code\cite{Pizzi2020}. It was carefully checked that all obtained MLWF are real-valued and reproduce the DFT band structure very accurately. The obtained Wannier functions are very localized with maximal spreads of \SI{2.18}{\angstrom\squared} for valence WF and \SI{5.25}{\angstrom\squared} for conduction WF. Since the underlying DFT-GGA calculations do not provide the correct band gap, we apply a scissors shift of \SI{0.9}{\eV} which is is similar to previously calculated quasi-particle shifts\cite{Bechstedt2015}. The Wannier Hamiltonians for valence and conduction bands provide all single-particle contributions of the exciton Hamiltonian Eq.~\eqref{eq:single_particle_H_wannier}.
The two-particle integrals entering $\tilde{H}^\text{SC}$ and $\tilde{H}^\text{LFE}$ are evaluated on a regular grid in Fourier space as described in Section~\ref{sec:evaluating_coulomb} and \ref{sec:evaluating_lfe}, which captures a supercell of $11\times11\times11$ primitive unit cells. The grid is determined by the Fourier space grid of the \textsc{vasp} calculation. (Overlap-)densities and auxiliary functions are also constructed on this real space grid and Fourier transformations (c.f. Eqs.~\eqref{eq:FT_fc},~\eqref{eq:FT_fv} and \eqref{eq:FT_lfe}) are performed using the FFTW library\cite{Frigo2005}. For the screening model introduced in Sect. \ref{section:general_formalism} we use $\epsilon_\infty=11.68$ for Si.
From the obtained single-particle and two-particle contributions we construct the exciton Hamiltonian Eq.~\eqref{eq:exciton_Hamiltonian_wannier} in a sparse matrix format where $\vS,\vS'$ are running over 61 lattice vectors in each direction for converged results. To test the capability of the LSWO approach we also performed calculations with 111 lattice vectors in each direction, which is equivalent to 1.37 million $\vk$-points.

The time evolution for the calculation of $\epsilon^\text{M}$ (c.f. Section~\ref{sec:time_domain_approach}) is performed by a Chebyshev polynomial expansion \cite{Weisse2006, Fan2021} of the time evolution operator, which has proven to be very accurate and efficient in the past\cite{Panhans2020,Panhans2021,Merkel2022}.
We set the maximum time to $\SI{14.77}{\pico\second}$, use $2000$ time steps and $16$ polynomials.
When calculating the spectrum we assumed a broadening of $\eta=\SI{65}{\milli\eV}$. Fig.~\ref{fig:autocorrelation} shows the time-autocorrelation function which enters Eq.~\eqref{eq:time_approach}.

We also carefully tested the implementation of the LSWO approach at multiple levels. This includes the comparison to an analytic Wannier-Mott exciton model and the reproduction of its energies. The interested reader is referred to Section~\ref{sec:implementation_test} of the appendix for more details.

\section{Results}

\subsection{Overlap densities and Coulomb integrals}

\begin{figure}
	\includegraphics[width=\textwidth]{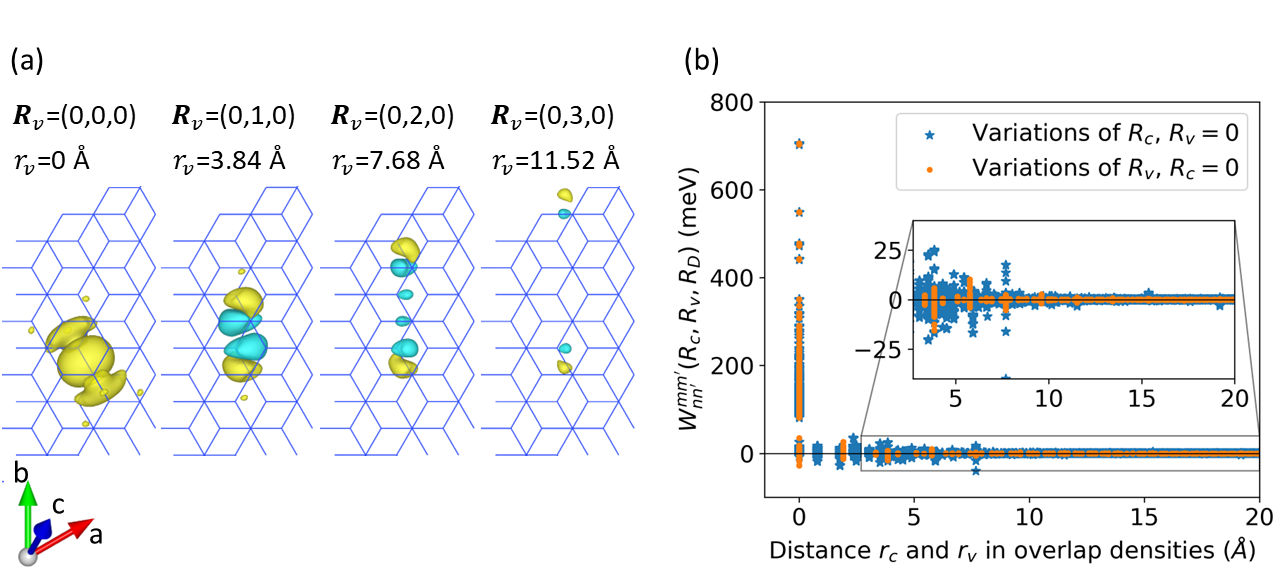}
	\caption{(a): Examples of overlap densities $\rho_{nn'\vR_v}$ for valence WF with $n=n'$ and different $\vR_v$. Yellow colours represent positive and blue negative values. All densities are plotted for the same iso-value magnitude of 0.001 and blue lines indicate the Si crystal.
	(b): Coulomb integrals for different hole-hole distances $r_v$ and electron-electron distances $r_c$ in the corresponding overlap densities $\rho_{nn'\vR_v}$ and $\rho_{mm'\vR_c}$. While $\vR_v$ and $\vR_c$ are only unit cell vectors, $r_v$ and $r_c$ also consider the position of Wannier centres within their unit cell.} \label{fig:sparse}
\end{figure}

Before discussing the optical absorption of bulk Si, we investigate more closely the distance-dependence of the two-particle contributions of the exciton Hamiltonian. We start by discussing the overlap densities $\rho_{mm'\vR_c}(\vx)$ and $\rho_{nn'\vR_v}(\vx)$, which contribute to the screened Coulomb interaction via Eq.~\eqref{eq:Coulomb_integral}. Fig.~\ref{fig:sparse}(a) shows selected overlap densities $\rho_{nn'\vR_v}$ of the valence WF (with $n=n'$ and different $\vR_v$). In this case, the overlap density for $\vR_v=0$ is a classical charge density in the shape of $\sigma$-bonded combination of $sp^3$ hybrid orbitals. The density is positive everywhere (yellow colour) with total charge of one. On the other hand, finite shifts $\vR_v$ introduce negative regions (blue colour) in $\rho_{nn\vR_v}$ and result in a total charge of zero. It is clearly seen that large values of $\vR_v$ lead to smaller overlaps as expected.

The implications of the decay of the Coulomb integrals $W^{mm'}_{nn'}(\vR_c,\vR_v,\vR_D)$ with distance are shown in Fig.~\ref{fig:sparse}(b). Blue stars denote data with varying distance between conduction WF $r_c$ and orange dots show data with varying distance between valence WF $r_v$. The distances $r_c$ and $r_v$ depend on the unit cell separation $\vR_c$ and $\vR_v$, respectively, and on the position of the Wannier centres within the unit cell.
It is clearly visible that already small separations in the overlap densities of a few angstroms lead to much smaller values in the Coulomb integral. The largest Coulomb integrals are observed for $r_c = r_v = 0$, where classical charge densities (with total charge of one) interact with each other. Our above discussion has therefore been confirmed numerically.
Furthermore, $W^{mm'}_{nn'}(\vR_c,\vR_v,\vR_D)$ is more sensitive to $r_v$ than $r_c$ because valence WFs are more localized than conduction WFs. In both cases, the overlap densities $\rho_{mm'\vR_c}$ and $\rho_{nn'\vR_v}$ vanish for large separations where the Coulomb integrals become zero. As a consequence, the corresponding screened Coulomb operator $\tilde{H}^\text{SC}$ is very sparse and the largest values contribute to the diagonal of the Hamiltonian matrix, as suggested. Similar results can be found for $\tilde{H}^\text{LFE}$ (not shown), which leads to a very sparse total exciton Hamiltonian.

\begin{figure}
	\centering
	\includegraphics[width=0.5\textwidth]{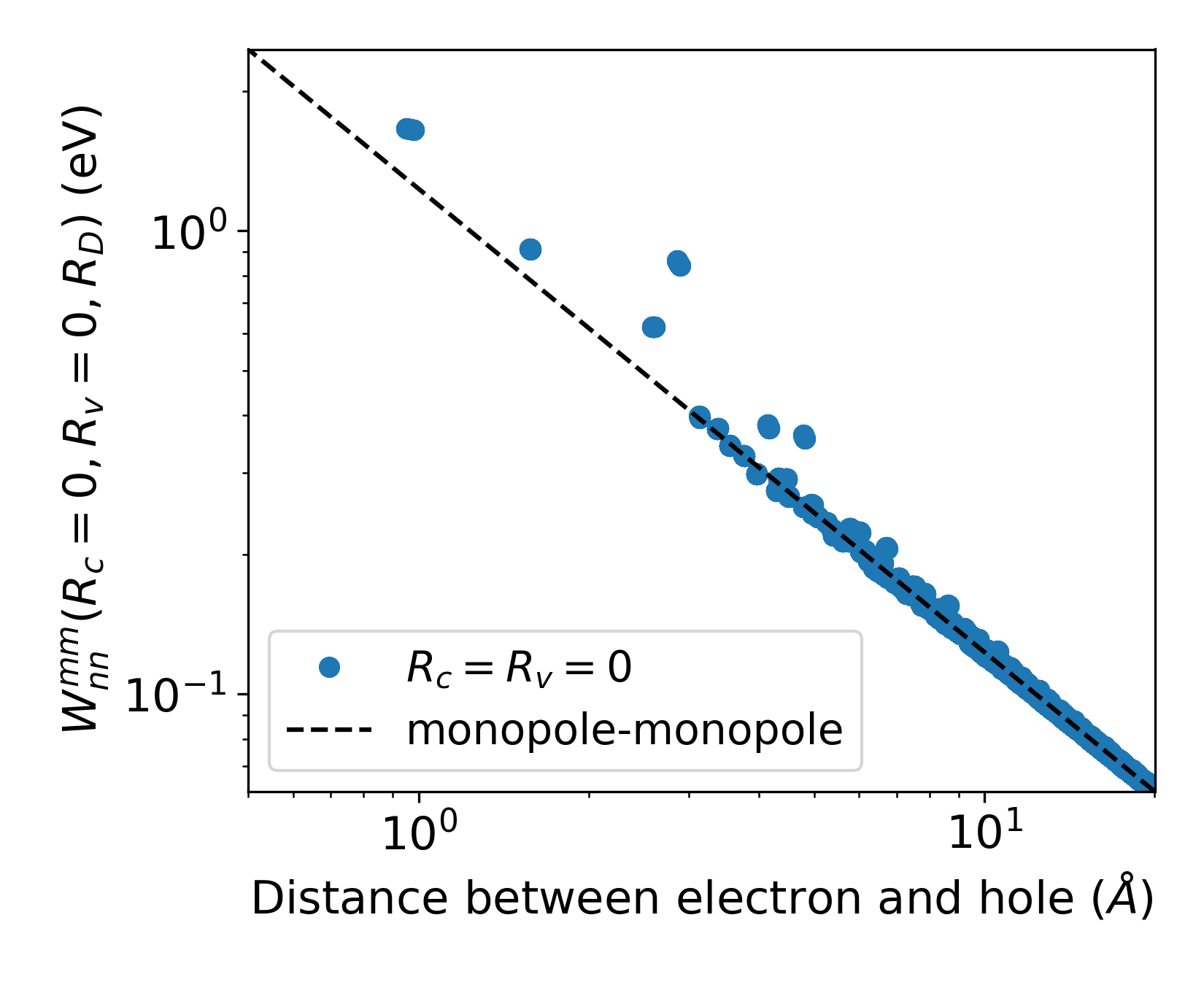}
	\caption{Screened density-density Coulomb interaction ($m=m'$, $n=n'$, $\vR_c=\vR_v=0$) between conduction and valence WF. The interaction is dominated by the monopole-monopole interaction (dashed line). Only interactions between overlapping densities with small distances differ significantly.} \label{fig:coulomb}
\end{figure}

We next turn to the diagonal elements of the Hamiltonian that correspond to electron-hole interaction of classical charge densities. They are shown in Fig.~\ref{fig:coulomb} for different distances between electrons and holes, which depends on the unit cell distance $\vR_D$ and the positions of the MLWF (charge centres) within a unit cell.
The Coulomb integrals $W^{mm}_{nn}(0,0,\vR_D)$ become smaller with increasing distance and can be approximated for distances larger than \SI{10}{\angstrom} by the monopole-monopole interaction (grey dashed line).
Notable deviations from the monopole-monopole approximation are found here only when electron and hole densities start overlapping at smaller distances. 
 As a result of the multipole expansion, only a relatively small fraction of the Coulomb integrals need to be calculated numerically, which reduces the computational effort substantially. 
For example, in the present study, we only need to compute 2496 out of 5.4 million density-density Coulomb integrals in full detail (less than \SI{0.5}{\percent} for a $61\times61\times61$ supercell with 4 valence and 6 conduction WFs) and assume the monopole-monopole approximation for the vast majority of terms. 
In general, the value of \SI{10}{\angstrom} does not have to be universal and deviations from the leading monopole-monopole term could occur also at larger distances, for instance in systems with Wannier functions that are less strongly localized. However, we are confident that systems with larger orbital spreads can also be treated very efficiently.

\subsection{Optical absorption spectrum}

\begin{figure}[h]
	\centering
	\includegraphics[width=0.8\textwidth]{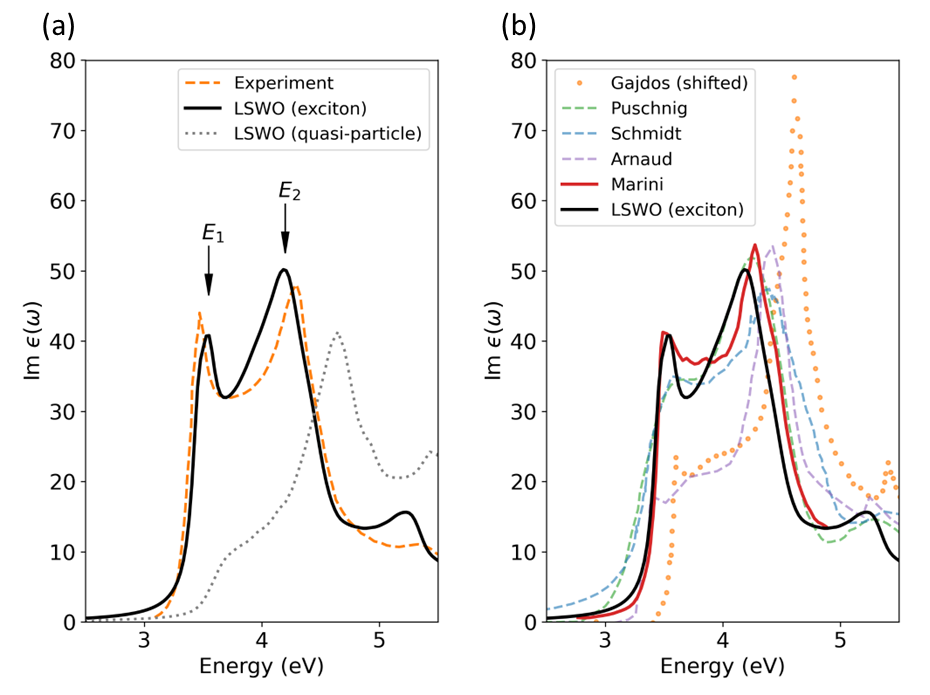}
		\caption{Absorption spectrum for silicon. (a): Comparison of the calculated MLWF-based spectrum (solid black) with calculations without electron-hole interaction (dotted grey) and experiment\cite{Lautenschlager1987} (dashed orange). Peak labels are in agreement with previous conventions\cite{Cardona2005}.
	(b): Comparison with other theoretical calculations. References: Gajdos\cite{Gajdos2006}, Puschnig\cite{Puschnig2002}, Schmidt\cite{Schmidt2003}, Arnaud\cite{Arnaud2001}, Marini\cite{Marini2008}} \label{fig:comparison_absorption}
\end{figure}

With the obtained exciton Hamiltonian we calculate the optical absorption spectrum of Si. Fig.~\ref{fig:comparison_absorption}(a) shows a comparison the LSWO approach (black solid line) to experimental data (orange dashed line). The spectrum contains the peaks $E_1$ and $E_2$ (naming convention from Ref.~\cite{Cardona2005}), in good agreement with experiment. Most importantly, the characteristic (direct) exciton peak at $E_1=\SI{3.5}{\eV}$ is a clear sign of bound exciton states that arise from electron-hole interactions. This peak is not present at GW or DFT theory level as shown by the dotted gray line. Compared to the quasiparticle spectrum, the excitonic effects result in a significantly redshifted spectrum, as generally expected which is a consequence of the electron-hole interaction. Residual deviations of the exciton spectrum to experiment might be related to the screening model (which is frequently used but still remains an approximation) or missing quasi-particle corrections in the band structure that go beyond a scissors shift.
Fig.~\ref{fig:comparison_absorption}(b) compares LSWO results to other theoretical calculations. The height of the $E_1$ exciton peak varies significantly among different methods, which might be related to different treatments of the screening.  
Our results are closely comparable to the approach by Marini\cite{Marini2008} and performs better than others in the literature.

\subsection{Scaling and performance of the LSWO approach} \label{sec:scaling}

\begin{figure}[h]
	\centering
	\includegraphics[width=0.9\textwidth]{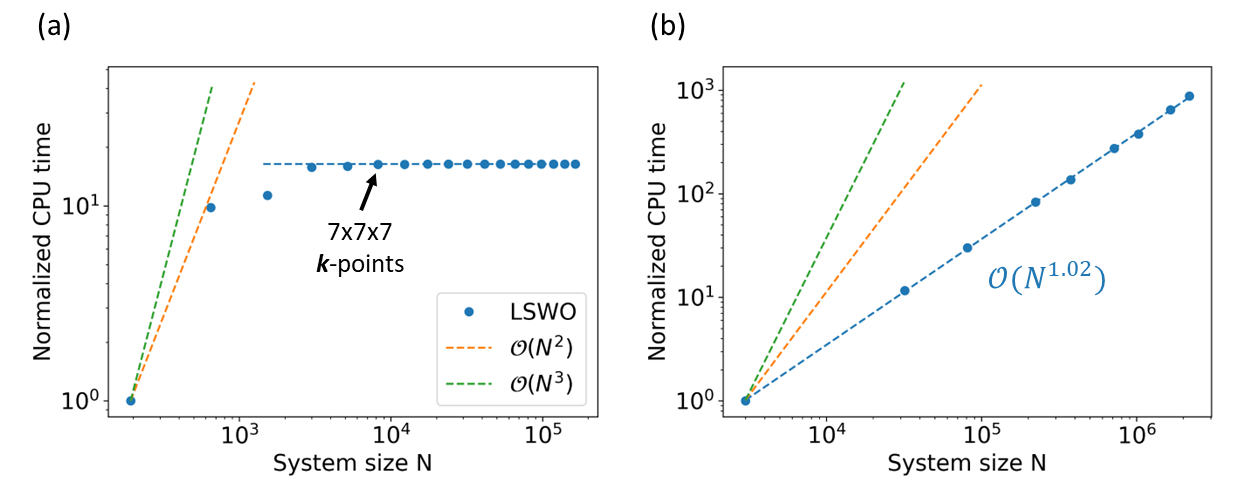}
	\caption{Scaling behaviour for (a) construction of the exciton Hamiltonian and (b) calculation of the optical absorption spectrum. $N$ is the rank of the Hamiltonian $N=N_\text{el}\cdot N_\text{h}\cdot N_{\vk}$. For comparison, a direct diagonalization of the exciton Hamilton in the Bloch basis (dense matrix) scales with $\mathcal{O}(N^3)$. Using the time evolution approach of Ref.~\cite{Schmidt2003} scales with $\mathcal{O}(N^2)$. The legend is shared for both figures. Calculations are performed on a single CPU core.} \label{fig:scaling}
\end{figure}

Finally, we discuss the performance and scaling with respect to the size of the exciton Hamiltonian, which depends on the number of valence and conduction states and the number of $\vk$-points (or equivalently $\vS$-points in Eq.~\eqref{eq:excition_hamiltonian_wannier}).
The overall performance depends on two parts, i.e., firstly the calculation of all required matrix elements of the Hamiltonian and secondly the evaluation of the optical absorption spectrum using the time evolution approach. Fig.~\ref{fig:scaling} shows the scaling of both parts for various numbers of $\vk$-points. All computations are performed on a single CPU core and normalized to a reference computation. Note that in the current implementation we do not exploit the symmetry of the crystal.

The most time-consuming part for the construction of the exciton Hamiltonian, which is shown in Fig.~\ref{fig:scaling}(a), is the evaluation of the Coulomb and LFE integrals that enter $\hat{H}^\text{SC}$ and $\hat{H}^\text{LFE}$. In contrast, the time required to generate the single-particle contributions of the Hamiltonian, i.e. valence and conduction bands, is negligible. As a result, the computing time scales with the number of two-particle integrals that need to be evaluated numerically on a real space grid. As we have shown in the previous section, the majority of such integrals either vanish if $\vR_c$ or $\vR_v$ deviate sufficiently from zero, or become analytical monopole-monopole interactions for larger values of $\vR_D$. Consequently, only a finite number of integrals need to be evaluated, leading to a saturation of CPU time in Fig.~\ref{fig:scaling}(a). This plateau is already reached for a supercell of $7\times7\times7$-unit cells (corresponding to a $\vk$-lattice of the same dimensions) which can be done with moderate effort. Once all integrals have been obtained, one can proceed to even denser $\vk$-grids (corresponding to very large supercells $\vS$) without additional effort for the computation of $\tilde{H}$.

The second step that is crucial to the performance of the LSWO method is the time evolution with the exciton Hamiltonian, which is shown in Fig.~\ref{fig:scaling}(b). This time propagation is performed in a step-by-step fashion, where each time step has the computational complexity of a sparse matrix-vector multiplication. Such operations can be performed very efficiently in linear scaling as shown in the figure. For comparison, the time-evolution approach in a Bloch representation, where the Hamiltonian is dense, would scale with $\mathcal{O}(N^2)$ \cite{Schmidt2003}, which is similar to implementations that use a Lanczos-Haydock approach as implemented in the Yambo code \cite{Alliati2022}. Note that a direct diagonalization of the Hamiltonian scales with $\mathcal{O}(N^2)$ in the case of a sparse matrix or with $\mathcal{O}(N^3)$ in the case of a dense matrix.

\section{Conclusion and outlook}
We have presented a method for describing the exciton Hamiltonian of the Bethe-Salpeter equation using maximally localized Wannier functions, which represent a minimal, spatially localized and material-specific basis set that accurately represents the quasiparticle band structure. 
The electron-hole interaction, i.e., local field effects and screened Coulomb attraction, are evaluated numerically in this basis, where the required number of two-particle matrix elements to be computed is greatly reduced due to the localized character of Wannier functions. Moreover, Coulomb integrals where electron and hole densities have large distances can be treated very efficiently in monopole approximation. 
Therefore this description in real space leads to a very sparse exciton Hamiltonian that can be calculated and used with high efficiency and offers intuitive user control over the simulations. With this implementation at hand, the macroscopic dielectric function for optical properties is calculated in the time domain using a linear-scaling algorithm. We have demonstrated the approach for a Si crystal where the optical subspace was constructed with millions of simple unit cells (corresponding to millions of $\vk$-points). The calculated absorption spectrum agrees well with experimental results.

In the future, we expect that the described LSWO approach will be very efficient for materials with many atoms per unit cell, which are not accessible with alternative current implementations. We hope that excitonic effects in optical spectra, which are relevant in a large number of crystalline systems, become more easily accessible. 

\section{Data availability statement}
The data that support the findings of this study are available in this article, the appendix or upon reasonable request from the authors.

\appendix
\renewcommand\thefigure{S-\arabic{figure}}
\setcounter{figure}{0}

\section{Step-by-step derivation for screened Coulomb interaction}

We inserting Eq.~\eqref{eq:xi_wannier} into $\tilde{H}^\text{SC}$ and using the shifting property of Wannier functions, i.e. $w_{m\vR}(\vx) = w_{m0}(\vx-\vR)$,
\begin{align}
&\tilde{H}^\text{SC}_{mn \vS,\, m'n' \vS'} = \int dx \int dx' \, \xi_{mn\vS}(\vx,\vx')   W(\vx-\vx') \xi_{m'n'\vS'}(\vx,\vx') \nonumber\\
&= \frac{1}{N_\Omega} \sum_{\vR\vR'}\int dx \int dx' \, 
 w_{m\vR}(\vx) w_{m'\vR'}(\vx)   W(\vx-\vx')  w_{n',\vR'-\vS'}(\vx') w_{n,\vR-\vS}(\vx') \nonumber\\
&= \frac{1}{N_\Omega} \int dx \int dx' \, 
\sum_{\vR\vR'} w_{m0}(\vx-\vR) w_{m'\vR'}(\vx)   W(\vx-\vx')  w_{n',\vR'-\vS'}(\vx') w_{n,\vR-\vS}(\vx') \nonumber\\
&= \frac{1}{N_\Omega} \sum_{\vR\vR'} \int dx \int dx' \, 
 w_{m0}(\vx) w_{m'\vR'}(\vx+\vR)   W(\vx+\vR-\vx')  w_{n',\vR'-\vS'}(\vx') w_{n,\vR-\vS}(\vx') \nonumber\\
&= \frac{1}{N_\Omega}\sum_{\vR\vR'} \int dx \int dx' \, 
 w_{m0}(\vx) w_{m'\vR'-\vR}(\vx)   W(\vx-\vx')  w_{n',\vR'-\vS'}(\vx'+\vR) w_{n,\vR-\vS}(\vx'+\vR) \nonumber\\
&= \frac{1}{N_\Omega}\sum_{\vR\vR'} \int dx \int dx' \, 
 w_{m0}(\vx) w_{m'\vR'-\vR}(\vx)   W(\vx-\vx')  w_{n',\vR'-\vR-\vS'}(\vx') w_{n,-\vS}(\vx') \nonumber\\
&= \frac{1}{N_\Omega} \int dx \int dx' \, 
\sum_{\vA\vB} w_{m0}(\vx) w_{m'\vA}(\vx)   W(\vx-\vx')  w_{n',\vA-\vS'}(\vx') w_{n,-\vS}(\vx') \nonumber\\
&= \sum_{\vA} \int dx \int dx' \, 
w_{m0}(\vx) w_{m'\vA}(\vx)   W(\vx-\vx')  w_{n',\vA-\vS'}(\vx') w_{n,-\vS}(\vx')
\end{align}
with $\vA=\vR'-\vR$ and $\vB=\vR'+\vR$.

An alternative form can be derived easily,
\begin{align}
&\tilde{H}^\text{SC}_{mn \vS, \,m'n' \vS'} = \nonumber \\
&= \sum_{\vA} \int dx \int dx' \, 
w_{m0}(\vx) w_{m'\vA}(\vx)   W(\vx-\vx')  w_{n',\vA-\vS'}(\vx') w_{n,-\vS}(\vx') \nonumber\\
&= \sum_{\vA} \int dx \int dx' \, 
w_{m0}(\vx) w_{m'\vA}(\vx)   W(\vx-\vx')  w_{n',\vA-\vS'}(\vx') w_{n,0}(\vx'+\vS) \nonumber\\
&= \sum_{\vA} \int dx \int dx' \, 
w_{m0}(\vx) w_{m'\vA}(\vx)   W(\vx-(\vx'-\vS))  w_{n',\vA-\vS'}(\vx'-\vS) w_{n,0}(\vx') \nonumber\\
&= \sum_{\vA} \int dx \int dx' \, 
w_{m0}(\vx) w_{m'\vA}(\vx)   W(\vx-\vx'+\vS)  w_{n',\vA+\vS-\vS'}(\vx') w_{n,0}(\vx') \nonumber \\
&= \sum_{\vA} \tilde{W}^{mm'}_{nn'}(\vA, \vS, \vS')
\end{align}

We finally show that the hermiticity relation of the Hamiltonian can be traced back to relations between single Coulomb integrals $\tilde{W}^{mm'}_{nn'}(\vA, \vS, \vS')$. For this we substitute $\vA\to-\vA$.
\begin{align}
\tilde{W}^{mm'}_{nn'}&(-\vA, \vS, \vS') = \nonumber\\
&= \int dx \int dx' \, 
w_{m0}(\vx) w_{m'-\vA}(\vx)   W(\vx-\vx'+\vS)  w_{n',-\vA+\vS-\vS'}(\vx') w_{n,0}(\vx')  \nonumber\\
&= \int dx \int dx' \, 
w_{m0}(\vx) w_{m'0}(\vx+\vA)   W(\vx-\vx'+\vS)  w_{n'0}(\vx'+\vA-\vS+\vS') w_{n,0}(\vx')  \nonumber\\
&= \int dx \int dx' \, 
w_{m0}(\vx-\vA) w_{m'0}(\vx)   W(\vx-\vA-(\vx'-\vA+\vS-\vS')+\vS) \times\nonumber\\
&\times w_{n'0}(\vx') w_{n,0}(\vx'-\vA+\vS-\vS')  \nonumber\\
&= \int dx \int dx' \,
w_{m\vA}(\vx) w_{m'0}(\vx)   W(\vx-\vx'+\vS')  w_{n'0}(\vx') w_{n,\vA-\vS+\vS'}(\vx')  \nonumber\\
&= \tilde{W}^{m'm}_{n'n}(\vA, \vS', \vS)
\end{align}
Performing the sum over $\vA$ on both sides, we obtain the hermiticity relation of the Hamiltonian.

\section{Model screening potential} \label{sec:model_screening_potential}

We start from the screened potential Coulomb interaction as defined in Section~\ref{section:general_formalism} and define $\alpha' = \alpha /q_\text{TF}^2$ for simplicity,
\begin{align}
W(\vq) = \epsilon^{-1}(\vq) V(\vq) &= \left(1- \frac{1}{\eta + \alpha' q^2}\right) \frac{1}{\epsilon_{0} q^2}.
\end{align}
A simple rearrangement of the terms yield the Coulomb and Yukawa potential in reciprocal space,
\begin{align}
W(\vq) &= \left(1- \frac{1}{\eta} \right)\frac{1}{\epsilon_{0} q^2} + \left(\frac{1}{\eta} -  \frac{1}{\eta + \alpha' q^2}\right) \frac{1}{\epsilon_{0} q^2} \nonumber\\
&= \frac{1}{\epsilon_\infty}\frac{1}{\epsilon_{0} q^2} + \frac{\alpha' q^2}{\eta(\eta + \alpha' q^2)} \frac{1}{\epsilon_{0} q^2} \nonumber\\
&= \frac{1}{\epsilon_{0} \epsilon_\infty q^2} + \frac{1}{\eta \epsilon_{0}}\frac{\alpha' }{\eta + \alpha' q^2} \nonumber\\
&= \underbrace{\frac{1}{\epsilon_{0} \epsilon_\infty q^2}}_{\text{= Coulomb}} + \left(1- \epsilon_\infty^{-1}\right) \underbrace{\frac{1}{\epsilon_{0}}\frac{1}{q^2 + \frac{q_\text{TF}^2}{\alpha(1- \epsilon_\infty^{-1})}}}_{\text{=Yukawa}}. \label{eq:model_screenning_reciprocalspace}
\end{align}
The Fourier transform then yields Eq.~\eqref{eq:model_screenning_realspace}.

\section{Numerical details}

\begin{figure}[h]
	\centering
	\includegraphics[width=0.6\textwidth]{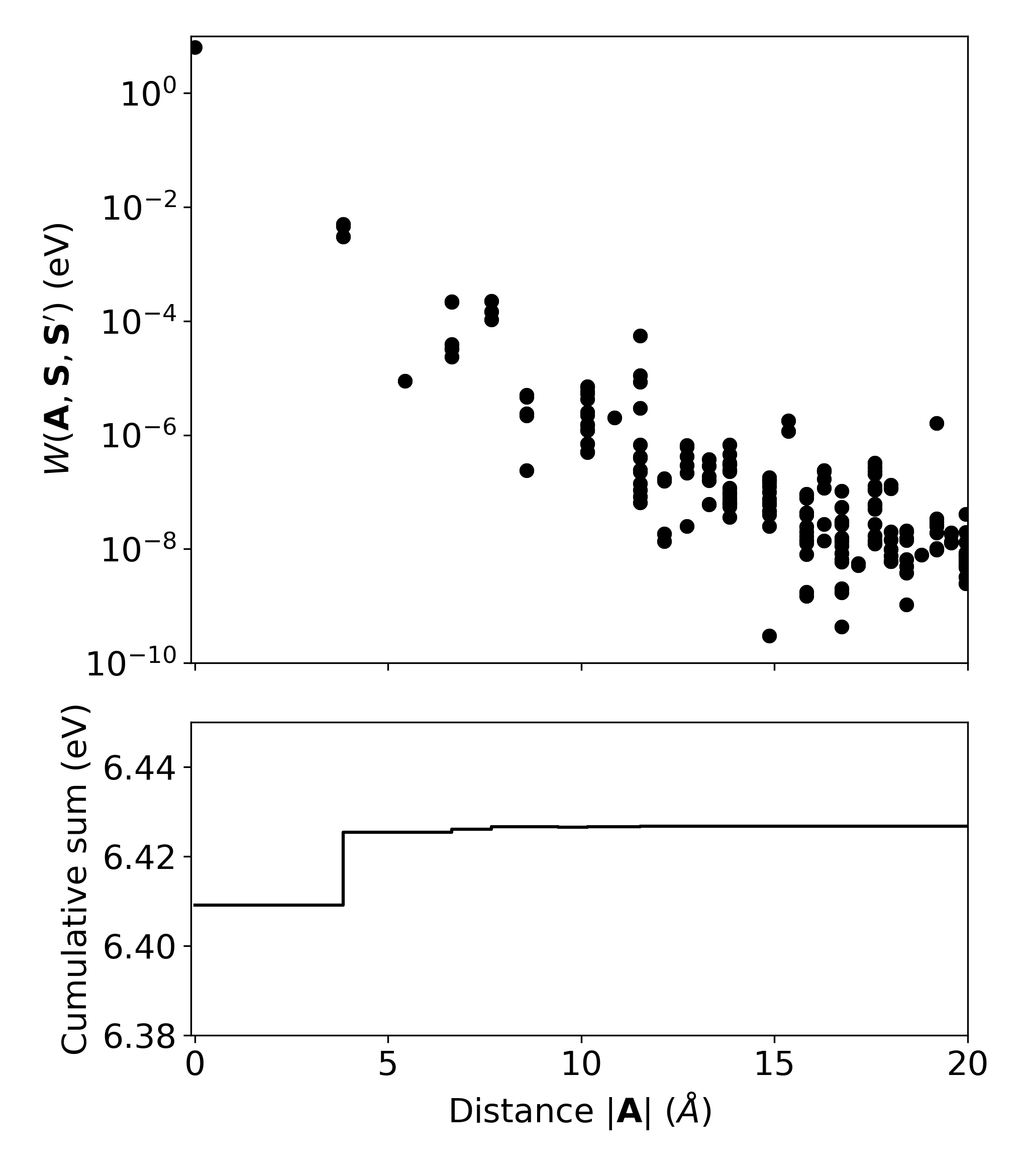}
	\caption{Convergence of Eq.~\eqref{eq:Coulomb_integral_wannier} for matrix element $\vS=\vS'=0$, $m=m'=1$, $n=n'=1$ }
\end{figure}

\begin{figure}[h!]
	\centering
	\includegraphics[width=0.6\textwidth]{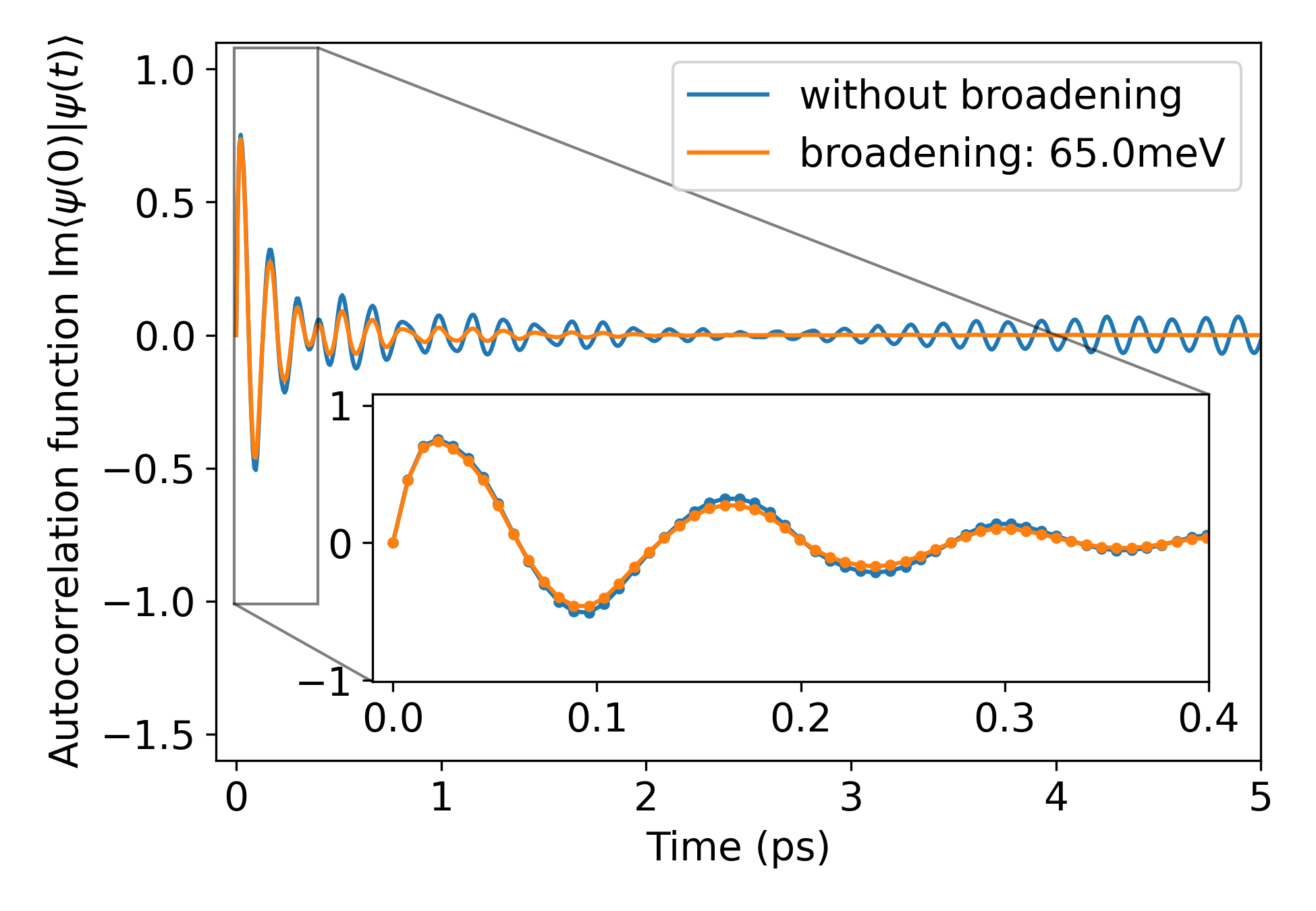}
	\caption{Imaginary part of the autocorrelation function that enters Eq.~\eqref{eq:time_approach} and is calculated using Eq.~\eqref{eq:time_evolution}. The initial state $\ket{\psi(t=0)}$ is normalized for convenience. } \label{fig:autocorrelation}
\end{figure}
\newpage

\subsection{Implementation test} \label{sec:implementation_test}

We have carefully and extensively tested all implementations, of which we want to discuss one particular test case that demonstrates the ability to compute excitons. For this purpose, we propose a simple test system that can be solved analytically. It consists of one orbital per unit cell in a cubic lattice of length $L$ and nearest neighbor transfer integrals for electrons and holes. The electronic structure is given by a tight-binding model,
\begin{align} \label{eq:hydrogen_tb}
H_\text{el} &= \sum_{<ij>} -t_\text{el}\,\, a_i^\dagger a_j + E_0, \nonumber \\
H_\text{h} &= \sum_{<ij>} t_\text{h} \,\,h_i^\dagger h_j,
\end{align}
whose band energies are
\begin{align}
E_\text{el}(\vk) &= -2t_\text{el} \left( \cos(k_xL) +\cos(k_yL)+\cos(k_zL)\right) + E_0, \\
E_\text{h}(\vk) &= 2t_\text{h} \left( \cos(k_xL) +\cos(k_yL)+\cos(k_zL)\right).
\end{align}

We construct the exciton Hamiltonian and include the electron-hole interaction. For simplicity we chose a static screening with $\epsilon_{\infty}$ and do not include local field effects. The resulting model is given by
\begin{align}
H(\vk,\vk') = \left[ E_\text{el}(\vk) - E_\text{h}(\vk) \right] \delta_{\vk\vk'} - \frac{1}{\epsilon_{\infty}} \tilde{V}(\vk-\vk'),
\end{align}
where $\tilde{V}(\vk-\vk')$ is the bare Coulomb potential in $\vk$-space. The model system is therefore similar to the Wannier-Mott exciton model \cite{Wannier1937}.
To obtain an analytical solution of this model, we perform a Taylor expansion of the band energies around $\vk=0$
\begin{align} \label{eq:hydrogen_taylor}
E_\text{el}(\vk) - E_\text{h}(\vk) 
\approx& E_0 -2(t_\text{el} + t_\text{h}) \left( 3-\frac{1}{2}L^2|\vk|^2  + \frac{1}{24}L^4 |\vk|^4 - ...  \right)
\end{align}
By expanding the exciton Hamiltonian up to second order we obtain the hydrogen-like problem,
\begin{align}
H(\vk\vk') 
=& \frac{\hbar^2\vk^2}{2\mu} \delta_{\vk\vk'} - \frac{1}{\epsilon_{\infty}} \tilde{V}(\vk-\vk') + E_\text{g},
\end{align}
with an effective mass $\mu = \frac{\hbar^2}{2(t_\text{el} + t_\text{h})L^2}$ and $E_\text{g} = E_0 -6(t_\text{el} + t_\text{h})$ the band gap without electron-hole interaction. The exciton energies follow a Rydberg series,
\begin{align}
E_{n} =& E_\text{g} - \frac{R_\text{ex}}{n^2\epsilon_{\infty}^2}m
\end{align}
where the exciton Rydberg energy $R_\text{ex}$ and exciton Bohr radius $a_\text{B}$ are,
\begin{align}
R_\text{ex} =& \frac{e^4 \mu}{2(4\pi \epsilon_{0})^2 \hbar^2}, \nonumber \\
a_\text{B} =& \frac{4\pi\epsilon_{0} \epsilon_{\infty} \hbar^2 }{\mu e^2}.
\end{align}

We note that this result can be further improved by calculating the energy shifts due to the $k^4$ term in Eq.\eqref{eq:hydrogen_taylor}, which would correspond to a relativistic correction of the hydrogen atom (fine structure without spin-orbit coupling). In complete analogy, they can be calculated using perturbation theory (more details on the derivation can be found in Ref.~\cite{Griffiths.2012}),
\begin{align}
\Delta E_{nl}
=& -\frac{1}{12} \frac{E_n^2}{(t_\text{el} + t_\text{h})}  \left[ \frac{4n}{(l+1/2)} - 3 \right].
\end{align}

\begin{figure}[h]
	\centering
	\includegraphics[width=0.6\textwidth]{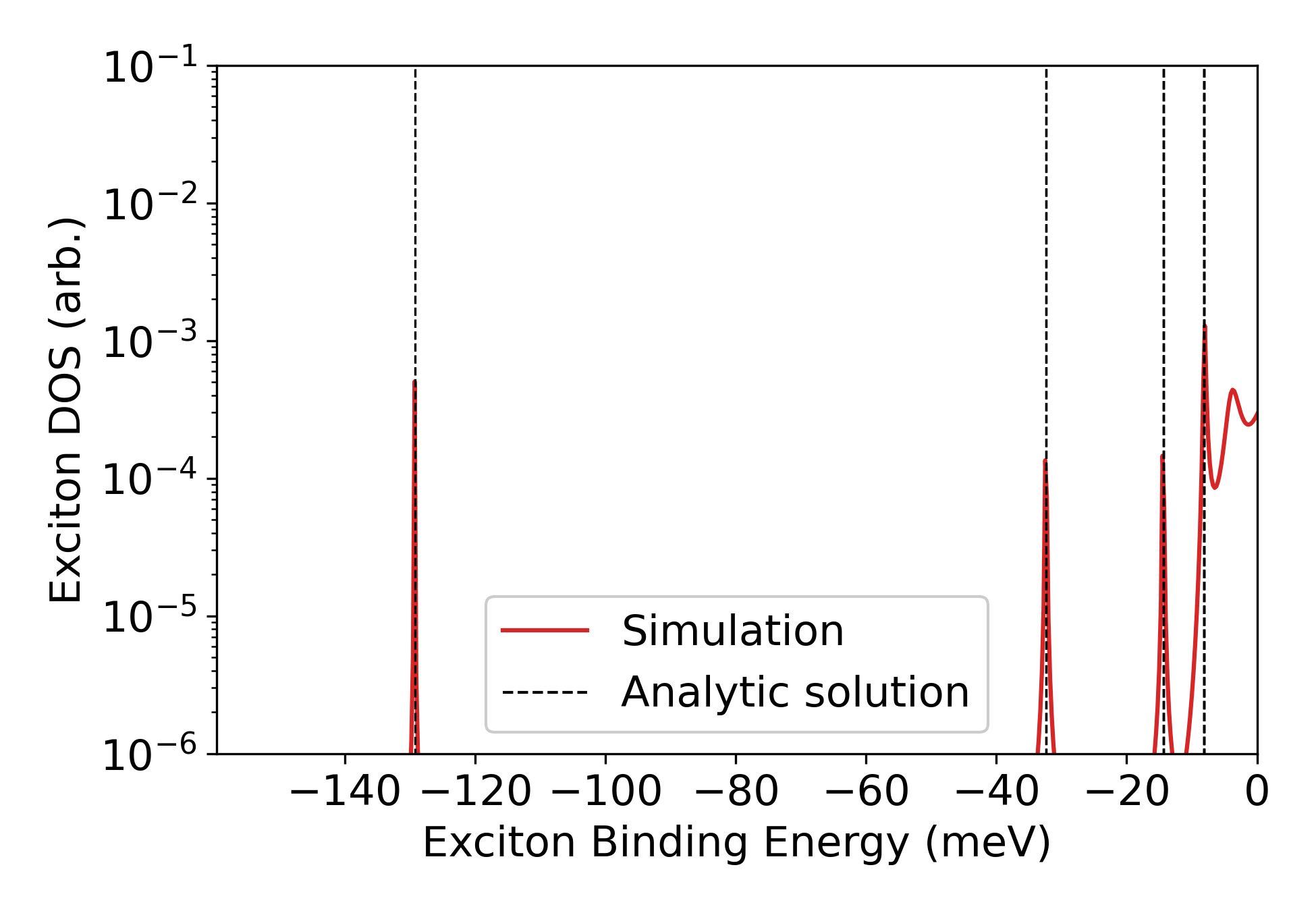}
	\caption{Rydberg series of Wannier excitons as a test case. Analytical results are shown as vertical dashed lines.} \label{fig:H-test}
\end{figure}

The analytical model will be compared with results of our Wannier implementation. Towards this end, the exciton Hamiltonian is set up in real space using the tight-binding models for valence and conduction bands (c.f. Eq.~\eqref{eq:hydrogen_tb}) and statically screened monopole-monopole interaction. The results can then compared for various model parameters ($L$, $t_\text{el}$, $t_\text{h}$ or $\epsilon_{\infty}$).
For converged numerical results, it is necessary to ensure that the size of the supercell (corresponding to the number of $\vk$ points) is large enough to host the eigenfunctions (hydrogen-like wavefunctions). More specifically, it must be much larger than the exciton Bohr radius $a_\text{B}$. To avoid discretization errors, the spacing of the lattice points must be small compared to $a_\text{B}$ so that the eigenfunction can be represented on a real space lattice.
By varying the parameters, one can obtain converged numerical results that are arbitrary close to the analytical result. On example is shown in Fig.~\ref{fig:H-test}, where the parameters are $L=\SI{5}{\angstrom}$, $t_\text{el} = t_\text{h} = \SI{8}{\eV}$, and $\epsilon_{\infty}=1$. The calculations are performed in a $700\times700\times700$ supercell and we have used an efficient Lanczos algorithm to calculate the density of states (DOS). The figure shows perfect agreement between the numerical and analytical results, demonstrating the correctness of our implementation and the ability to simulate various excitons.

\newpage

\section{ACKNOWLEDGEMENTS}
We would like to thank the Deutsche Forschungsgemeinschaft for financial support [CRC1415, projects No. OR-349/3 and OR-349/11 and the Cluster of Excellence e-conversion (Grant No. EXC2089)]. Grants for computer time from the Zentrum für Informationsdienste und Hochleistungsrechnen of TU Dresden and the Leibniz Supercomputing Centre in Garching (SuperMUC-NG) are gratefully acknowledged.

We would like to acknowledge F. Bechstedt and J. Furthmüller for fruitful discussions about the numerical evaluation of Coulomb integrals.

\section{COMPETING INTERESTS}
There are no competing interests to declare.


\end{document}